\documentclass[reprint, showpacs, superscriptaddress, aps, pre, twocolumn, 10pt]{revtex4-1}

\usepackage{graphicx} 
\usepackage{dcolumn}
\usepackage{bm}
\usepackage{amssymb}
\usepackage{amsmath}
\usepackage{wasysym}
\usepackage{color}
\usepackage{hyperref}
\usepackage{dcolumn}
\usepackage{float}
\usepackage{flushend}
\usepackage{multirow}
\usepackage{cancel}
\hypersetup{
colorlinks,
citecolor=blue,
filecolor=blue,
linkcolor=blue,
urlcolor=blue}

\newcommand{\be}{\begin{equation}}
\newcommand{\ee}{\end{equation}}
\newcommand{\bse}{\begin{subequations}}
\newcommand{\ese}{\end{subequations}}
\newcommand{\bea}{\begin{eqnarray}}
\newcommand{\eea}{\end{eqnarray}}
\newcommand{\ba}{\begin{array}}
\newcommand{\ea}{\end{array}}

\begin{document}
\title{Asymmetric energy transfers in driven nonequilibrium  systems and arrow of time}
\author{Mahendra K. Verma}
\email{mkv@iitk.ac.in}
\affiliation {Department of Physics, Indian Institute of Technology, Kanpur, India 208016}

\begin{abstract}
Fundamental  interactions are either fully or nearly symmetric under time reversal.  But macroscopic phenomena  have a definite arrow of time. Though there is no convergence on the origin of time's preferential direction, many researchers believe that the direction of time is towards increasing entropy.  In this paper, we provide an alternate point of view.  In driven-dissipative nonequilibrium systems forced at large scale, the energy flows from large scales to dissipative scales.  This generic and multiscale process breaks time reversal symmetry and principle of detailed balance, thus can yield an arrow of time.    In this paper we propose that conversion of large-scale coherence to  small-scales incoherence could be treated as a dissipation mechanism for generic physical systems.  We illustrate the above processes using turbulence as an example.  We argue that the above picture of time irreversibility could be employed to the universe and to many-body quantum systems. 

\end{abstract}

\maketitle

\section{Introduction}
Fundametal forces---gravity, electrodynamics, and strong nuclear---exhibit time reversal symmetry.   Weak nuclear force however exhibits a small violation of this symmetry.  Hence, from the perspectives of fundamental interactions, forward and backward motion of a physical system are indistinguishable  apart from a small violation for weak nuclear force~\cite{Feynman:book:Character,Lebowitz:RMP1999,Carroll:book:Time}. This observation contradicts daily experience where physical and biological systems evolve forward in time.  For example, stars, planets, and living beings take birth, live, and then die. There have been many attempts to explain this asymmetry, some of which will be summarized below. In this paper, we show how asymmetric energy transfers determine the arrow of time in driven-dissipative nonequilibrium systems.  We illustrate the properties of energy transfers for hydrodynamic  turbulence, which is a well-studied example of  driven-dissipative nonequilibrium system.

There is a general consensus that the arrow of time in physical systems arises due to the second law of thermodynamics, according to which  a nonequilibrium system evolves in such a way that the entropy of the system always increases~\cite{Feynman:book:Character,Lebowitz:RMP1999,Carroll:book:Time}.  For example, when we mix a set of cold molecules with another set of hot molecules, the mixture tends towards  a uniform distribution of molecules with maximum entropy.    The other mechanisms invoked for explaining arrow of time are sensitivity to initial condition in nonlinear systems~\cite{Strogatz:book},  measurements in quantum systems~\cite{Wheeler:book:Quantum_measurements}, chaos induced decoherence~\cite{Berry:book_chapter:QM}, etc.  The energy transfer mechanism presented in this paper is an alternative framework to break the time reversal symmetry.

Some of the leading examples of driven-dissipative nonequilibrium systems are turbulence, earthquakes, crack propagation,  fragmentation,  free market economy, astrophysical flows, etc.  Several common features among them are (a) energy supply at large scales; (b) energy cascade from large scale to intermediate scale and then to small scale; (c) dissipation at small scales; (d) multiscale physics with energy transfers across scales~\cite{Kolmogorov:DANS1941Dissipation,Kolmogorov:DANS1941Structure,Frisch:book}.    Each of the above systems are covered in vast literature.  Here, we take turbulence as an illustrative example because it has been widely studied, and it is familiar to  physicists.   

Unidirectional energy transfers (from large scales to small scales) break the time reversal symmetry in turbulence.  In a time-reversed version of a turbulent flow, the energy will flow from small scales to large scales, contrary to real systems.  Jucha {\em et al.}~\cite{Jucha:PRL2014} made the above observation, and then went on to quantify the irreversibility in turbulence using the relative motion between two particles. They showed that the difference between the forward and backward dispersion of the two particles varies as $t^3$, where $t$ is the time elapsed. Davidson~\cite{Davidson:book:Turbulence} attributed the arrow of time in a turbulent flow to chaotic advection arising due to the nonlinear terms.  

In the present paper we focus on the energy transfers in  turbulence and show how they play a critical role in determining the arrow of time.  In turbulence, the energy transfers  break the principle of detailed balance due to asymmetric energy transfers from large scales to small scales.  We show  that the above scenario is applicable to many nonequilibrium systems.  Also,  we argue that the dissipation of kinetic  energy at small scales and its subsequent conversion to heat offers an interesting recipe for introducing dissipation in a generic multiscale system.     In the paper we also emphasise that the above multiscale description involving energy transfers  differs significantly from  the prescription of second law of thermodynamics.    Towards the end of the paper we  argue how the ideas of energy transfers could be extended to the universe and  many-body quantum systems.

The outline of the paper is as follows: In Sec.~\ref{sec:turbulence}, we describe how asymmetric energy transfers in turbulence  can set the direction of time.  In Sec.~\ref{sec:dissipation}  we argue that the conversion of large-scale coherent energy to small-scale incoherent energy could be treated as a generic dissipation mechanism in nonequilibrium systems.  We conclude in Sec.~\ref{sec:conclusions}.

\section{Direction of time in turbulent systems}
\label{sec:turbulence}

In this section we  describe the essential physics of turbulence, and then the arrow of time in such flows.    Consider an incompressible hydrodynamic flow that is forced at large length scales  ($L_f$, forcing scale).  According to the celebrated Kolmogorov's theory~\cite{Kolmogorov:DANS1941Dissipation,Kolmogorov:DANS1941Structure}, in a turbulent flow, the energy supplied at large scales cascades to intermediate scales (called {\em inertial range}) and then to dissipative scales.  See Fig.~1a for an illustration.  Note that the  energy supply rate, the energy flux in the inertial range, and the energy dissipation rate are all equal (denoted by $\epsilon_u$)~\cite{Kolmogorov:DANS1941Dissipation,Landau:book:Fluid,Frisch:book,Lesieur:book:Turbulence}. 

\begin{figure}
\centering
\includegraphics[width=1\linewidth]{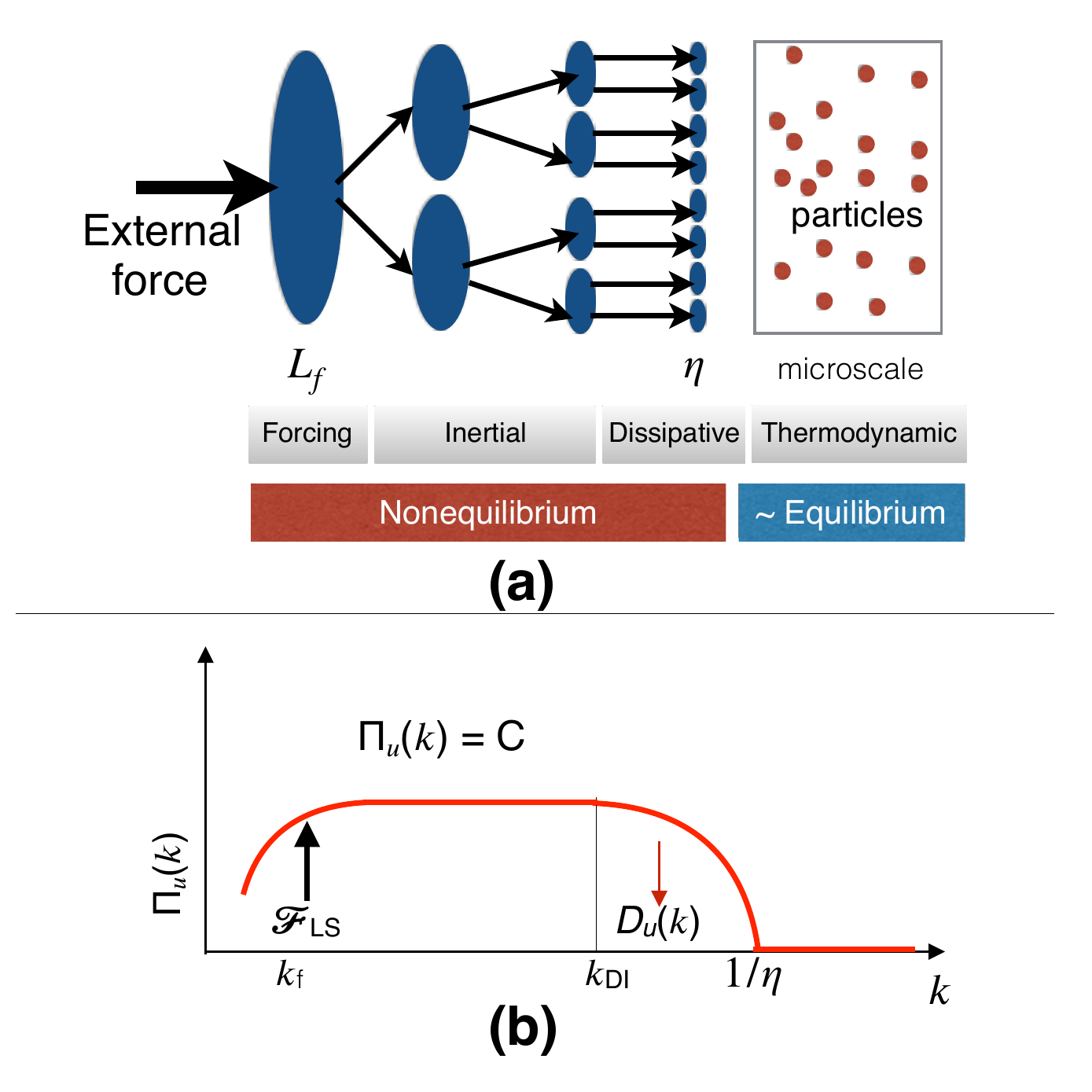}
\caption{(color online) Schematic diagrams illustrating energy transfers in three-dimensional hydrodynamic turbulence, which is an example of a driven-dissipative  nonequilibrium system:  (a) The energy supplied at large scales cascades to the inertial range and then to the dissipative range.  The dissipation of coherent kinetic energy  at the dissipative scales  heats up the particles at microscale.  The energy transfers are indicated by arrows.  The structures at the forcing and inertial range are governed by nonequilibrium processes, while those at microscale are in approximate equilibrium.  (b) Plot of spectral energy flux $\Pi_u(k)$ vs.~$k$.   $\Pi_u(k) \sim \mathrm{const}$ in the inertial range, and it decreases with $k$ in the dissipative range.  $\mathcal{F}_\mathrm{LS}$ is the energy supply by the external force, and $D_u(k)$ is the dissipation rate.  There is no energy flux at the microscales due to the detailed balance. }
\label{fig:NS}
\end{figure}

In the above macroscopic picture, which is based on {\em  continuum approximation}, the velocity field at a position ${\bf r}$ is averaged over many microscopic particles around ${\bf r}$.  This description of the velocity field is sensible up to Kolmogorov length scale $\eta$, which is approximately equal to $(\nu/\epsilon_u^3)^{1/4}$, where $\nu$ is the kinematic viscosity of the fluid.  The microscopic physics, e.g. kinetic theory, describes the dynamics of particles at length scales smaller than $\eta$ because the particles at these scales  are under an approximate equilibrium.  However, the system is dynamic and in a nonequlibrium state for $l > \eta$ (see Fig.~1a). Thus, the length scale $\eta$ plays a major role in separating the coherent hydrodynamic structures with random motion at microscopic scales.   We illustrate these features using an example.  When we pour milk in a coffee cup, the large fluid structures of milk successively breaks  into a hierarchy of  structures, until it diffuses at microscales where the macroscopic coherent energy is destroyed as disorder or heat. 

It is customary to describe turbulence in Fourier space too with wavenumber $k \sim 1/l$.     In this description, as shown in Fig.~1b, the energy flux (or energy cascade rate) $\Pi_u(k) \sim \mathrm{const}$  in the inertial range, and it dampens in the dissipative range $k_\mathrm{DI} < k < 1/\eta$.  There is no energy flux in the microscopic range. 

The nonlinear term of the Navier-Stokes equation, ${\bf u \cdot \nabla u}$ where ${\bf u}$ is the velocity field, induces energy transfers from one scale to another scale. In Fourier space, the basic unit of nonlinear interactions is a wavenumber triad $({\bf k, p,q})$ satisfying ${\bf k=p+q}$.    Dar {\em et al.}~\cite{Dar:PD2001} and Verma~\cite{Verma:PR2004} showed that  the energy transfer from Fourier mode ${\bf u(p)}$ to Fourier mode ${\bf u(k)}$ under the mediations of Fourier mode ${\bf u(q)}$ is given by
\be
S^{uu}({\bf k|p|q}) = \Im \left[ {\bf  \{  k \cdot u(q) \} \{ u(p) \cdot u^*(k) \} } \right] .
\label{eq:ET:Suu_kpq}
\ee
Using the above function we define the energy flux $\Pi_u(k_{0})$, which is  the energy emanating from a wavenumber sphere of radius $k_0$ due to nonlinear interactions, as:
\begin{equation}
\Pi_u(k_{0})=\sum_{|\mathbf{p}|\le k_{0}} \sum_{|\mathbf{k}|>k_{0}} S^{uu}({\bf k|p|q}) .
\label{eq:fluid_flux}
\end{equation}
 A nonzero flux implies that the phases of the Fourier modes are coherently organised.  The flux would be zero if the phases of the Fourier modes are random and uncorrelated.   Kolmogorov~\cite{Kolmogorov:DANS1941Dissipation,Kolmogorov:DANS1941Structure} and Pao~\cite{Pao:PF1968}  showed that  the energy spectrum of the inertial-dissipation range is given by
\bea
E_u(k) & \approx &  K_\mathrm{Ko} \epsilon_u^{2/3} k^{-5/3} \exp\left(-\frac{3}{2}  K_\mathrm{Ko} (k\eta)^{4/3}\right),  \label{eq:Ek}\\
\Pi_u(k) & \approx & \epsilon_u  \exp\left(-\frac{3}{2}  K_\mathrm{Ko} (k\eta)^{4/3}\right),
\label{eq:Pik}
\eea
where $k = |{\bf k}|$, and $K_\mathrm{Ko}$ is Kolmogorov's constant whose value is approximately 1.6~\cite{Frisch:book}.  The aforementioned Kolmogorov's spectrum and flux have been verified by many experiments and numerical simulations.  For a recent numerical work, refer to \citet{Verma:FD2018}.

Now we investigate in some detail the energy transfers and flux  in a three-dimensional turbulent flow.  It has been shown that for a turbulent flow~\cite{Verma:PR2004,Verma:Pramana2005S2S},
\bea
S^{uu}({\bf k|p|q}) > 0~\mathrm{for}~k > p, \\
S^{uu}({\bf k|p|q})  < 0~\mathrm{for}~k < p.
\eea
Hence,  energy flows preferentially from large scales to small scales, thus breaking the time reversal symmetry as well as the detailed balance (of energy transfer). The energy flux, which is a sum of $S^{uu}$'s [Eq.~(\ref{eq:Pik})], is positive in the inertial range.  These energy transfers  create and sustain a hierarchy of structures.  Note that in a time-reversed system, ${\bf u} \rightarrow {\bf -u}$, substitution of which in Eqs.~(\ref{eq:ET:Suu_kpq}, \ref{eq:fluid_flux}) leads to opposite $S^{uu}({\bf k|p|q})$ and negative $\Pi_u(k)$, implying  flow of energy from small scales   to large scales.  Clearly, such a process is not allowed from physical considerations (Kolmogorov's theory of turbulence), hence the time-reversed flow can be differentiated from the real flow.  Thus, the energy transfers and flux  set the arrow of time in a turbulent system.

 For a steady incompressible turbulent flow, the energy contents of structures at the large and intermediate scales remain unchanged on an average.   Hence the measure of disorder (entropy) is not expected to increase for these structures.  Thus, entropy is not very important for the description of steady-state turbulent structures of an incompressible flow.  However, the energy supplied at large scales is finally transferred to the molecules at microscales.  Hence the molecules at the microscopes get heated up, thus  increasing the entropy at  these scales.    We also remark that the dynamics (Navier-Stokes equations) and viscous dissipation play a key role in turbulence and driven-dissipative nonequilibrium systems.   Broadly speaking, the forcing and inertial range, as well as part of dissipation range, would be in nonequilibrium state; and a fraction of dissipation range and all of microscopic range would be in an approximate equilibrium state.  See Fig.~\ref{fig:NS} for an illustration.   We remark that turbulence does not violate second law of thermodynamics.  Here, we argue that the unidirectional energy transfers in turbulence provide an alternative viewpoint for time  reversal asymmetry. 

 The aforementioned transfers from large scales to small scales  are  observed in many driven-dissipative nonequilibriun systems---for examples in  earthquakes~\cite{Turcotte:book:Fractals}, magnetohydrodynamic turbulence~\cite{Verma:PR2004},  scalar turbulence and thermal convection~\cite{Yeung:JFM2013,Shraiman:Nature2000,Verma:book:BDF}, finance~\cite{Verma:book_chapter:2019}, and astrophysical and geophysical flows~\cite{Goldstein:ARAA1995,McWilliams:book:GFD}.   In thermal convection, large-scale thermal plumes drive the turbulence.   Galactic and stellar turbulence have behaviour similar to hydrodynamic turbulence with energy feed at large scale by supernovas and star core (where nuclear reaction takes place) respectively~\cite{Zeldovich:book:Magneticfield}.  Collisions of tectonic plates feed energy to earthquakes, and this energy is transmitted to smaller scales.  The money supply at large scales drives a free market economy. The energy transfers in the above systems differ in detail, but energy flows from large scales to small scales in all of them.  This crucial feature sets the arrow of time in these nonequilibrium systems.
 
Exceptions to the aforementioned picture are two-dimensional and quasi-two-dimensional (e.g. rotating) turbulence where the energy flows from small scales to large scales~\cite{Boffetta:ARFM2012,Davidson:book:Turbulence,Lesieur:book:Turbulence,Sharma:PF2018} due to an inverse cascade of energy.   These transfers lead to formation of large scale structures.  Such systems do not violate our model for the direction of time.  If our system is two-dimensional and quasi-two-dimensional turbulence, then we define the forward direction of time as one in which the energy transfers are from small scales to  large scale.  We have to carefully observe both the space dimensionality and energy transfers  for determining the direction of time.  

We remark that the viscous dissipation at small scales plays an important role in setting the direction of energy flow in driven-dissipative nonequilibrium systems.  Typically, the energy supplied at large scales flows towards the energy sink at the dissipative scale.      Given this, an  important question is how to introduce an effective dissipation in a nonequilibrium system.   Here too, the physics of turbulence gives importance clues, as described in the next section.

\section{How to incorporate dissipation in a physical system?}
\label{sec:dissipation}

Fundamental forces of nature are conservative or nondissipative.  Hence, inclusion of dissipation  appears impossible from the perspectives of fundamental physics.  Yet, multiscale description of turbulence  provides  interesting possibilities for  an inclusion of dissipation in  a physical system.

An important question: what is the origin of the viscous dissipation in a  fluid flow when it  is composed of many interacting molecules and atoms, as visualised in statistical mechanics?  Naively, we do not expect any dissipation in a flow.  The answer however becomes apparent when we think in terms of multiscale physics, and separate the coherent and incoherent structures.  As shown in Fig.~\ref{fig:NS}, in a turbulent flow, the relative velocity between the fluid structures causes successive cascade of coherent energy to smaller and smaller scales. Finally, the coherent structures are fully dissipated at Kolmogorov's microscale ($\eta$).  Hence, in hydrodynamic turbulence, viscous dissipation provides an interesting model for transferring energy from coherent structures to incoherent ones. We emphasize that in  a decaying turbulence (with no external force), the total kinetic energy of the flow, $\rho u^2/2$ where $\rho$ is the fluid density, decays.  But the total energy, which is a sum of kinetic energy of the fluid and particles,  is conserved.  Thus,  the viscosity converts coherent kinetic energy to thermal energy at the microscales.  Similar processes occur in plasma turbulence where the energy supplied from the large-scale velocity and magnetic fields are dissipated at microscales by kinetic processes involving charge particles~\cite{Elmegreen:ARAA2004,Verma:JGR1996viscosity}.  

 We expect a similar separation of scales in many driven-dissipative nonequilibrium systems---astrophysical turbulence (in stars, galaxies,  galaxy clusters, etc.),  free market economy with cascade of money, earthquakes, etc.     The energy (or similar quantity, e.g. money in a financial system) flows from large scales to small scales, where it is dissipated.   The physical processes at small scales that converts coherent energy to incoherent ones could be treated as dissipation.  Even the  frictional loss during the motion of a block on a surface can be treated as conversion of coherent kinetic energy of the block to the incoherent heat energy of molecules at the interface.   Introduction of dissipation in the universe  and many-body (macroscopic) quantum systems appears to be a possibility under this perspective.

The universe is very complex due to intricate   space-time structures,  complicated and enormous energy sources and sinks, unknown initial condition (during the early universe), etc.  Gravitational interactions are conservative. Yet, from hydrodynamic perspectives,  it is  possible to treat the universe as a driven-dissipative nonequilibrium system with energy  flowing from large scales (driven by sources like supernova) to intermediate scales and then to small scales. For example, large-scale kinetic energy of a galaxy is transferred to the motion of dust and/or charged particles.   Such mechanisms are employed for the generation of magnetic field in the universe~\cite{Zeldovich:book:Magneticfield}.  Note however that dynamical interactions among the entities could lead to formation or destruction of structures.  For example, depending on the forces, (local) space dimensionality, and initial configurations, the flow could become less structured or more structured.  As described in  Section~\ref{sec:turbulence},   large-scale structures are formed in two-dimensional or quasi-two-dimensional turbulence~\cite{Boffetta:ARFM2012,Davidson:book:Turbulence,Lesieur:book:Turbulence,Sharma:PF2018}.  Note that strong rotation~\cite{Sharma:PF2018}, as in galaxies, or buoyancy due to density-stratification~\cite{Verma:book:BDF}, as in planetary and stellar atmospheres, can make the flow quasi-two-dimensional.  Hence, the aforementioned dynamical and multiscale perspective with energy transfers  could possibly provide a  cosmological arrow of time  for the universe (even in the scenario of collapsing universe).  


Dissipation or decoherence in quantum systems is quite intriguing, and it remains primarily an unsolved problem.  Some of the proposed mechanisms for quantum decoherence are collapse of wave function during a measurement, interaction with enviornment~\cite{Wheeler:book:Quantum_measurements}, chaos~\cite{Berry:book_chapter:QM}, etc.  In this short article, we do not delve into these topics, but describe recent observations on dissipation in quantum turbulence.  Recent experiments and numerical simulations on superfluid turbulence allude to certain similarities between  quantum turbulence and classical Navier-Stokes turbulence~\cite{Nemirovskii:PR2012}.  For a wavenumber band, $E_u(k) \sim k^{-5/3}$ with a constant energy cascade.  Bradley {\em et al.}~\cite{Bradley:PRL2006} argued that energy dissipation in Helium-3 superfluid turbulence, which is quantum  many-body system, occurs via  phonon coupling.  Thus microscopic processes like phonon interactions in Helium-3 could provide dissipation of large-scale quantum correlations (also see \citep{Fonda:PNAS2019}).     It would be interesting to attempt similar ideas for other  many-body quantum systems, such as Bose-Einstein condensate, quantum  cavity with many atoms, etc.

Thus, a conversion of (cascaded) coherent energy to incoherent one at small scales provide an interesting way to incorporate dissipation in a driven-dissipative nonequilibrium system, as well as in many-body classical and quantum systems.  This idea however needs further explorations using experiments and numerical simulations.

\section{Conclusions}
\label{sec:conclusions}
In summary,  unidirectional and multiscale energy transfers in turbulence and in driven-dissipative nonequilibrium systems provide an alternate formulation for determining the arrow of time in such systems.     In this framework, dissipation at small scales and energy supply at large scales play a critical role.     An important outcome of such formalism is a recipe to introduce dissipation in multiscale systems.  We propose that the transformation from coherent energy to incoherent energy could be treated as dissipation in physical systems. 

Interestingly, many biological systems too  exhibit multiscale processes with energy supply at large scales,  sustenance at intermediate scale, and dissipation at small scales. For example, a living being receives food at large scales, which is used by organs (all the way to cells in a hierarchal manner) to generate energy and nutrition at intermediate scales.  The waste products (e.g. $CO_2$)  generated at small scales  are excreted out.  These time-asymmetric processes may be playing a key role in determining biological arrow of time.  

Before closing the discussion we remark that the mechanism for breaking time reversal symmetry based on unidirectional energy transfers is very different from those proposed in second law of thermodynamics, and in chaos theory.  According to the second law, an increase in entropy sets the direction of time.   Note that the energy transfer formalism described in this paper does not involve entropy computations of fluid structures.   Some researchers attribute the arrow of time in chaotic  systems  to {\em sensitivity to initial condition}.  In a chaotic  system, most   initial conditions  take the system to chaotic configurations with higher entropy, and only a small set of initial conditions evolve to  ordered states.  This  mechanism of chaotic dynamics differs significantly from that  involving energy transfers in a driven-dissipative nonequilibrium system.

 We hope that the aforementioned  multiscale framework may be useful for resolving some of the longstanding issues on the arrow of time in physical and biological systems, as well as in cosmology.

\acknowledgements
The author thanks Roshan Samuel, Anurag Gupta, Anand Jha, and Arul Lakshminarayan for useful  discussions.  

\end{document}